\documentclass{appolb}
\usepackage{graphicx}

\usepackage{amsmath} 
\usepackage{xcolor}
\usepackage[numbers]{natbib}
\usepackage{hyperref}

\definecolor{darkgreen}{rgb}{0,0.35,0.15}
\definecolor{Green4}{rgb}{0.,0.376,0.}
\definecolor{Gray}{rgb}{0.2,0.2,0.2}
\definecolor{Gray2}{rgb}{0.35,0.35,0.35}
\definecolor{Orange}{rgb}{0.9,0.4,0.}
\definecolor{Black}{rgb}{0.,0.,0.}
\definecolor{Blue}{rgb}{0.,0.,1.}
\definecolor{Green}{rgb}{0.,1.,0.}
\definecolor{Cyan}{rgb}{0.,1.,1.}
\definecolor{Red}{rgb}{1.,0.,0.}
\definecolor{Magenta}{rgb}{1.,0.,1.}
\definecolor{Yellow}{rgb}{1.,1.,0.}
\definecolor{White}{rgb}{1.,1.,1.}
\definecolor{Blue4}{rgb}{0.,0.,0.5625}
\definecolor{Blue3}{rgb}{0.,0.,0.6875}
\definecolor{Blue2}{rgb}{0.,0.,0.8125}
\definecolor{LtBlue}{rgb}{0.52734375,0.8046875,1.}
\definecolor{Green3}{rgb}{0.,0.6875,0.}
\definecolor{Green2}{rgb}{0.,0.8125,0.}
\definecolor{Cyan4}{rgb}{0.,0.5625,0.5625}
\definecolor{Cyan3}{rgb}{0.,0.6875,0.6875}
\definecolor{Cyan2}{rgb}{0.,0.8125,0.8125}
\definecolor{Red4}{rgb}{0.5625,0.,0.}
\definecolor{Red3}{rgb}{0.6875,0.,0.}
\definecolor{Red2}{rgb}{0.8125,0.,0.}
\definecolor{Magenta4}{rgb}{0.5625,0.,0.5625}
\definecolor{Magenta3}{rgb}{0.6875,0.,0.6875}
\definecolor{Magenta2}{rgb}{0.8125,0.,0.8125}
\definecolor{Brown4}{rgb}{0.5,0.1875,0.}
\definecolor{Brown3}{rgb}{0.625,0.25,0.}
\definecolor{Brown2}{rgb}{0.75,0.375,0.}
\definecolor{Pink4}{rgb}{1.,0.5,0.5}
\definecolor{Pink3}{rgb}{1.,0.625,0.625}
\definecolor{Pink2}{rgb}{1.,0.75,0.75}
\definecolor{Pink}{rgb}{1.,0.875,0.875}
\definecolor{Gold}{rgb}{1.,0.83984375,0.}

\definecolor{background}{cmyk}{0,0,0.3,0}
\definecolor{dgreen}{rgb}{0,.4,0}
\definecolor{plum}{rgb}{.7 .2 .7}
\definecolor{darkgreen}{rgb}{.0 .6 .0}
\definecolor{peru}{rgb}{.80 .52 .25}

\begin{document}
\title{The road to solving the Gribov problem of the\\ center vortex model in quantum chromo dynamics
\thanks{Presented at Excited QCD 2019 by Rudolf Golubich}
}
\author{Rudolf Golubich and Manfried Faber 
\address{Atominstitut, Technische Universit\"at Wien}
}
\maketitle
\begin{abstract}
The center vortex model of the QCD vacuum is very successful in explaining the non-perturbative properties of QCD, especially confinement, chiral symmetry breaking and the topological charge of vacuum configurations. On the other hand, the center vortex model still suffers from a Gribov problem: Direct maximal center gauge and center projection can lead to an underestimation of the string tension in smooth configurations or after persistent simulated annealing. We discuss methods to identify center regions, whose boundaries evaluate to center elements, and want to improve the vortex detection: these regions might help to recognize vortices in configurations where maximal center gauge lost the vortex finding property. 
\end{abstract}

\section{Introduction}
The \textit{center vortex model}~\cite{THOOFT,CORNWALL,DelDebbio:1998luz} is based upon the center symmetry of the action in lattice quantum chromo dynamics. It describes the properties of the vacuum by percolating vortices, which are closed and quantized magnetic flux lines of finite thickness, condensing in the vacuum. It is capable of explaining:
\begin{itemize}\setlength\itemsep{-1mm}
\item \textbf{Confinement:} behaviour of Wilson and Polyakov loops, see~\cite{DelDebbio:1998luz}
\item \textbf{Casimir scaling} of the potential due to thick vortices, see~\cite{Faber:1997rp}
\item \textbf{Broken scale invariance}, see~\cite{Langfeld:1997jx}
\item \textbf{Chiral symmetry breaking} can be linked to vortices in sufficiently smooth configurations~\cite{Hollwieser, Faber:2017alm}.
\end{itemize} 
But for some configurations it underestimates the string tension. As this might be due to failing vortex detection procedures, we intend to improve the methods of P-vortex detection, also striving towards a direct identification of thick vortices.
\section{The center vortex model}
\begin{figure}[h!]
\begin{center}
\includegraphics[width=10cm]{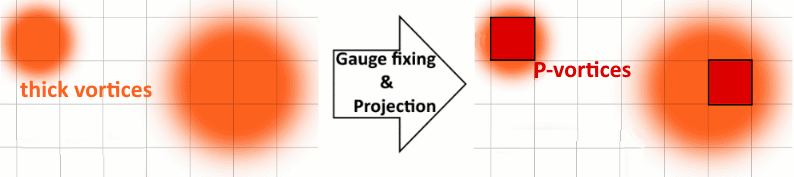}
\end{center}
\caption{Vortex finding property: P-vortices localize thick vortices.}\label{thickDetection}
\end{figure}

\begin{figure}[h!]
\parbox[c]{5cm}{
\includegraphics[width=5cm]{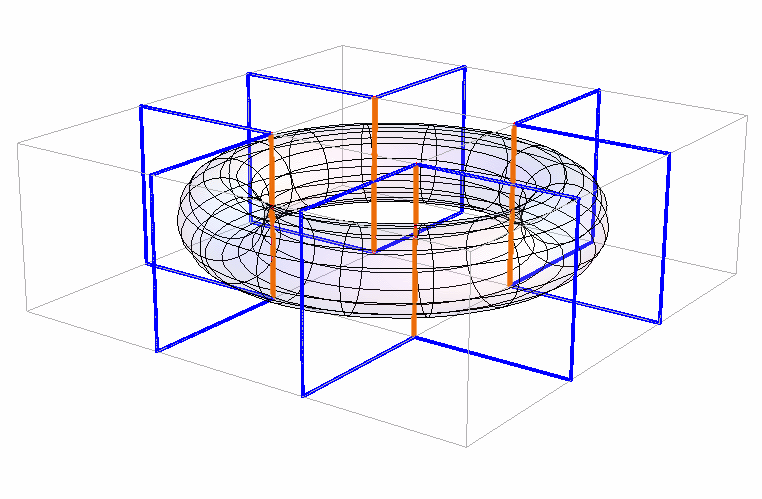}
} \parbox[c]{6cm}{
\textcolor{Orange}{Non-trivial links} build up the \textit{Dirac volume}, whose surface is the vortex (transparent). They are detected by  \textcolor{Blue2}{non-trivial plaquettes}.}
\caption{Relationship between center vortices, non-trivial links and the Dirac volume.}
\label{Vortex}
\end{figure}

Thick vortices are presently localized by P-vortices, which in turn are identified by maximal center gauge and center projection, see Fig.~\ref{thickDetection}.
Gauge fixing is done by maximizing the gauge fixing functional
\begin{equation}
R = \sum_x \sum_\mu \mid \text{Tr}[ U_{\mu}(x)] \mid^2,
\end{equation}
with $U_{\mu}(x)$ being the SU(2) link at position $x$ in direction $\mu$. Numerical methods like simulated annealing can only find local maxima. Some of the configurations corresponding to these maxima suffer from a loss of the vortex finding property. Projecting the single links to center elements
\begin{equation}
U_{\mu}(x) \rightarrow Z_{\mu}(x) = \text{sign } \text{Tr}[ U_{\mu}(x)].
\end{equation}
leads to P-vortices. They arise in the projected lattice as plaquettes evaluating to a non-trivial center element, see figure \ref{Vortex}.

\begin{figure}[h!]
\parbox[c]{4.5cm}{
\includegraphics[width=4.5cm]{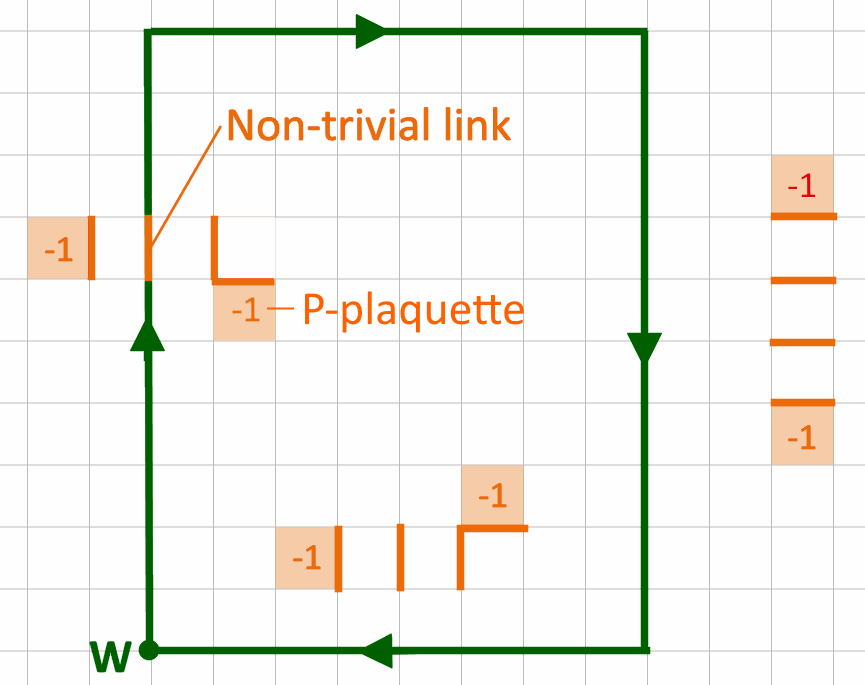}
}\vspace{0.5cm} \parbox[c]{6.5cm}{
Via the \textcolor{Orange}{Non-trivial links} of the Dirac volume, each P-vortex plaquette contributes a factor $-1$ to an enclosing \textcolor{Green4}{Wilson loop}. A loop enclosing an even number of P-plaquettes evaluates to the trivial center element while a loop enclosing an odd number of P-plaquettes evaluates to a non-trivial center element.
}
\caption{P-Vortices and Wilson loops.}
\label{Wilson}
\end{figure}

Assuming independent piercings of Wilson loop $W(R,T)$ of size $R\times T$ in the projected lattice, see figure \ref{Wilson}, follows an area law of the expectation value of Wilson loops,
\vspace*{-6mm}\begin{equation}
\langle \textcolor{Green4}{W(R,T)}\rangle = [\textcolor{Orange}{(-1)}\varrho+\textcolor{Blue}{(+1)}(1 - \varrho)]^{\overbrace{\textcolor{Green4}{R} \times \textcolor{Green4}{T}}^{A}} = e^{-\ln (1- 2 \varrho) A} = e^{-\sigma A}.
\end{equation}

This estimate relates the \textit{vortex density} $\varrho$, the percentage of P-vortex plaquettes, to the string tension $\sigma$
\begin{equation}
\sigma = \ln (1- 2 \varrho).
\end{equation}

\begin{figure}[h!]
\begin{center}
\parbox{82mm}{
\parbox[c]{30mm}{\includegraphics[scale=0.7]{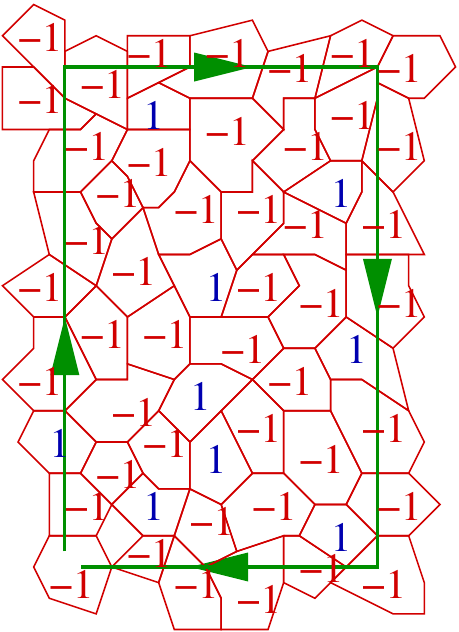}}
\parbox[c]{25mm}{$ \; = \;  \textcolor{Red2}{(-1)^{22}} \; \times$}
\parbox[c]{25mm}{\includegraphics[scale=0.7]{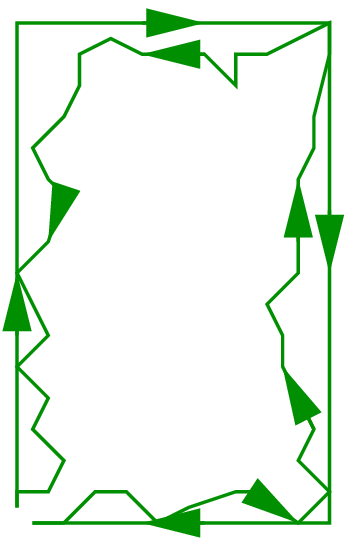}}
\\
\parbox{80mm}{
$ \; \underbrace{\qquad \qquad \qquad \qquad }_{\text{\large center regions}} \; \underbrace{ \quad \quad \quad \;}_{\text{\large Area law}} \quad \; \underbrace{\qquad \qquad \qquad}_{\text{ \large Perimeter law}}$}
} \hspace{2mm}\parbox[c]{3.8cm}{
Regions, whose boundaries evaluate to center elements can be used to factorize a Wilson loop into two parts: an area factor collecting the fully enclosed non-trivial regions, correlating to vortices, and a factor from non-center contributions, a perimeter factor.
}
\end{center}
\caption{Factorization of a Wilson loop using center regions.}
\label{Factorization}
\end{figure}

Due to the short range Coulomb field, the Creutz ratios
\begin{equation}
\chi(R,T)=\frac{\langle W(R+1,T+1) \rangle \; \langle W(R,T) \rangle}{\langle W(R,T+1) \rangle \; \langle W(R+1,T) \rangle}, \qquad \sigma \approx -\log \chi(R,R),
\end{equation}
of full configurations decrease with $R$ and approach an asymptotic value $\sigma_\infty$. 
\begin{figure}[h!]
\begin{center}
\includegraphics[height=4cm]{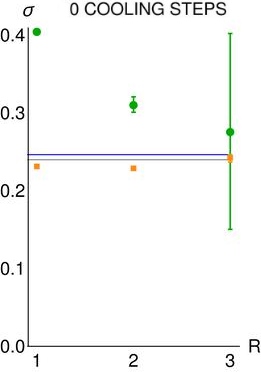}
\hspace{0.75cm}\includegraphics[height=4cm]{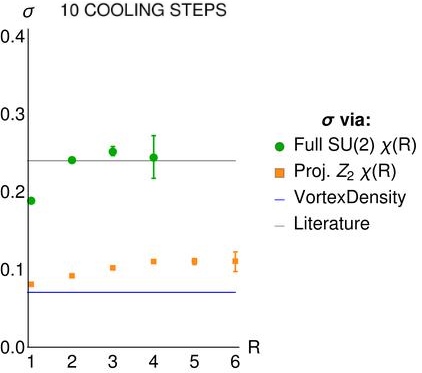}
\caption{String tension $\sigma$ calculated with different methods for $\beta=2.2 $ and lattice size $18^4$ with 460 SU(2) Wilson configurations. Left: Without cooling all calculations are compatible to the literature value. Right: With cooling (strength $0.05$) calculations based upon vortices underestimate the literature value.}
\label{Sigma}
\end{center}
\end{figure}
The corresponding asymptotic string tension is usually well reproduced by the center projected string tension, even for small $R$. For smooth configurations the center vortex model tends to underestimate $\sigma_\infty$, see Fig.~\ref{Sigma}. This might be due to a loss of the vortex finding property in some of the configurations of the ensemble. We suspected that this loss results in center regions having opposite sign in the projected $Z_2$ configurations \cite{goluFab}, see figure \ref{Factorization}, than in the full SU(2) configurations.

\section{Identifying center regions}
\begin{figure}[h!]
\includegraphics[width=12.25cm]{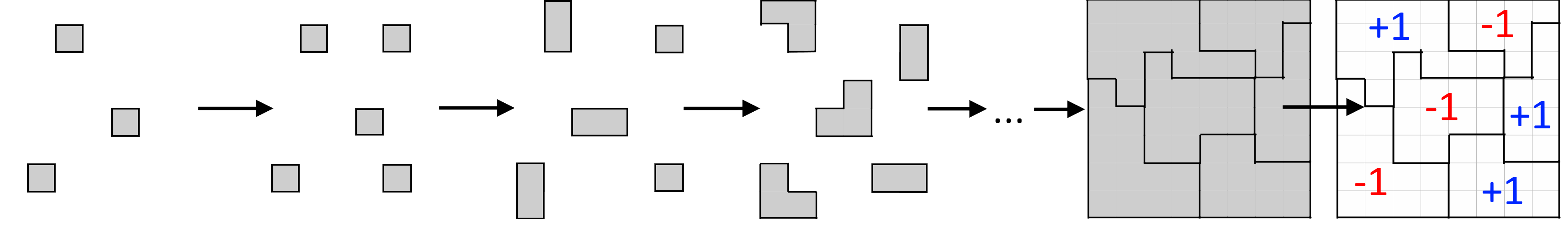}
\caption{P-Vortices and Wilson loops.}
\label{Algo1}
\end{figure}

The algorithm works down a stack of plaquettes sorted by rising trace. For each plaquette not already selected it tries to find a bigger region by adding a before unselected neighbouring plaquette whenever the combination of them results in a region with a more negative trace, see figure \ref{Algo1}.

\begin{figure}
\begin{center}
1)~\includegraphics[height=1.25cm]{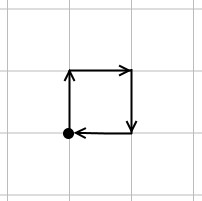}
~~~2)~\includegraphics[height=1.2cm]{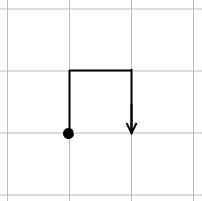}
~~~3)~\includegraphics[height=1.2cm]{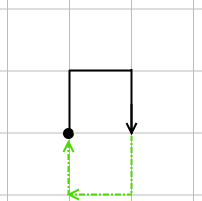}
~~~4)~\includegraphics[height=1.2cm]{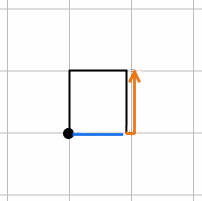}
~~~5)~\includegraphics[height=1.2cm]{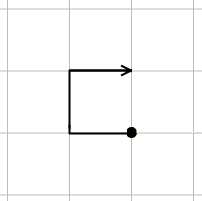}
\hspace{2cm}
\end{center}

\begin{center}
~~~6)~\includegraphics[height=1.2cm]{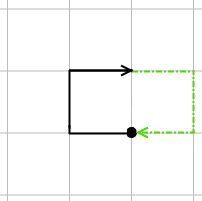}
~~~\textbf{...}~~~
~~~10)~\includegraphics[height=1.2cm]{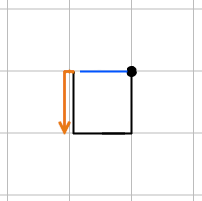}
~~~11)~\includegraphics[height=1.2cm]{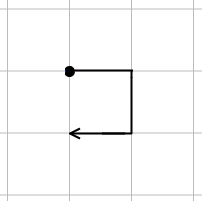}
~~~12)~\includegraphics[height=1.2cm]{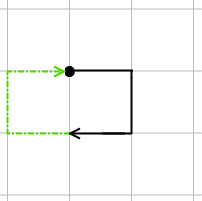}
~~~13)~\includegraphics[height=1.2cm]{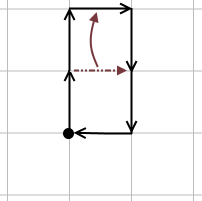}
\end{center}
\caption{Algorithm for enlarging regions.}
\label{Algo2}
\end{figure}

Figure \ref{Algo2} shows the enlargement procedure: A plaquette is stored as connected list of its links (1), an evaluation with one link missing is stored (2), by complementing this open region around the neighbouring plaquette the trace of a possible enlargement and the respective direction of enlargement are stored (3), by two multiplications the opening is moved along the perimeter of the region (4) and another possible complementation is calculated (6). If this complementation is better than the previously done, it overwrites the stored variables. This steps are repeated until the best direction of enlargement is identified and the connected list can be complemented around the respective neighbouring plaquette (13). 
This procedure only considers plaquettes that do not belong to already identified regions, hence it runs in linear time with respect to the number of plaquettes.

\section{Results}
\begin{figure}[h!]
\begin{center}
\includegraphics[height=5cm]{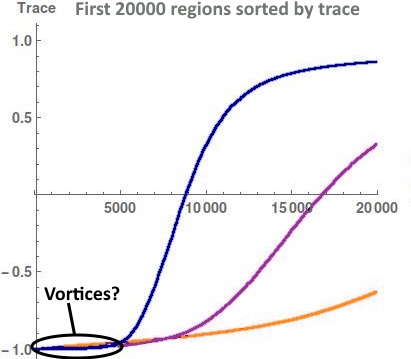}
\hspace{0.5cm}\includegraphics[height=4.5cm]{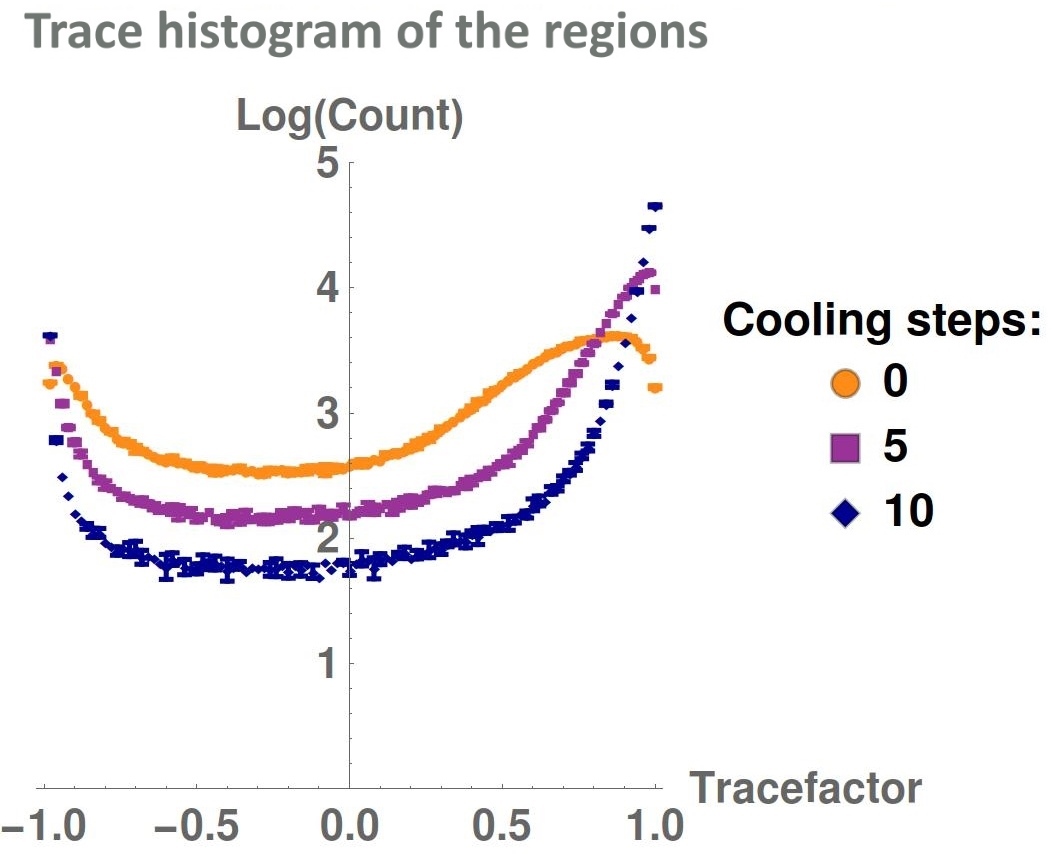}
\caption{Data taken from SU(2) Wilson configurations at $\beta=2.2 $ and lattice size $14^3\times8$ with cooling strength $0.005$. Left: The first 20000 regions sorted by rising trace, taken from one configuration. Right: Tracefactor-Histogram of the identified regions for different number of cooling steps taken from 5 configurations.}
\label{centerRegions}
\end{center}
\end{figure}

In figure \ref{centerRegions} the traces of the identified regions are depicted. 
After 10 cooling steps the algorithm still detects about 5000 non-trivial regions for $\beta=2.2$ and about 3000 for $\beta=2.4$. The correct reproduction of the string tension in the non-cooled configuration was based on $16874\pm401$ vortex plaquettes for $\beta=2.2$ and $7338\pm301$ for $\beta=2.4$.
After 10 cooling steps the maximal center gauge and center projection identify $4540\pm 249$ vortex plaquettes at $\beta=2.2$, but only $1534\pm165$ at $\beta=2.4$. Already with our preliminary algorithm for center region detection slight improvements of the vortex detection can be expected.

\bibliographystyle{unsrtnat}
\bibliography{literature}

\end{document}